\documentclass[twocolumn]{svjour3}

\smartqed

\usepackage[intlimits,fleqn]{amsmath}
\usepackage[mathscr]{eucal}
\usepackage[english]{babel}
\usepackage{graphicx}
\usepackage{txfonts}
\usepackage[latin1]{inputenc}

\journalname{Astrophysic and Space Science}

\begin{document}

\title{ALMA : Fourier phase analysis made possible}


\author{François Levrier \and Edith Falgarone \and François Viallefond}


\institute{F. Levrier \at
              LRA, ENS, 24 rue Lhomond 75231 Paris Cedex 05 \\
              \email{francois.levrier@ens.fr}
           \and
           E. Falgarone \at
              LRA, ENS, 24 rue Lhomond 75231 Paris Cedex 05 \\
              \email{edith.falgarone@ens.fr}
           \and
           F. Viallefond \at
              Observatoire de Paris, 61 Avenue de l'Observatoire, 75014 Paris \\
              \email{Francois.Viallefond@obspm.fr}
}

\date{Received: date / Accepted: date}

\maketitle

\begin{abstract}
Fourier phases contain a vast amount of information about structure in direct space, that most statistical tools never tap into. We address ALMA's ability to detect and recover this information, using the probability distribution function (PDF) of phase increments, and the related concepts of phase entropy and phase structure quantity. We show that ALMA, with its high dynamical range, is definitely needed to achieve significant detection of phase structure, and that it will do so even in the presence of a fair amount of atmospheric phase noise. We also show that ALMA should be able to recover the actual ``amount'' of phase structure in the noise-free case, if multiple configurations are used.

\keywords{Instrumentation: interferometers \and Methods: statistical \and Methods: numerical \and ISM: structure}
\end{abstract}

\section{Introduction}

Observations of the interstellar medium (ISM) reveal highly complex, fractal-like structures \cite{falgarone91,elmegreen96}. The self-similar hierarchy of these structures, over four decades, is thought to spring from the interplay of turbulent motions \cite{miesch94} and self-gravitation \cite{burkert2004}. To understand this interplay, one therefore needs a quantitative description of the observed structures. Most of the statistical tools used to this end are more or less derived from the power spectrum \cite{dickey2001}, which is given by the squared amplitudes of Fourier components. Yet, a simple numerical experiment performed by \cite{coles2005} shows that essential structural information lies in the Fourier-spatial distribution of the phases.

In the following, we present some of the notions used to exploit this information (section \ref{sec_pfa}), and their practical implementation (section \ref{sec_psqip}). We then consider the ability of ALMA and other arrays to detect and measure phase structure information in real time (section \ref{sec_atio}). We conclude by giving some future perspectives (section \ref{sec_conc}).

\section{Fourier phase analysis}
\label{sec_pfa}

The importance of Fourier phases in terms of structure has been recognized by various studies \cite{scherrer91,polygiannakis95,coles2005}. Since the information sought lies in the Fourier spatial distribution of phases, Scherrer {\it et al.} \cite{scherrer91} suggested considering the statistics of phase increments $\Delta_{\boldsymbol{\delta}}\phi(\boldsymbol{k})=\phi(\boldsymbol{k}+\boldsymbol{\delta})-\phi(\boldsymbol{k})$ between points separated by a given lag vector $\boldsymbol{\delta}$ in Fourier space. 

\begin{figure}[htbp]
\resizebox{0.85\hsize}{!}{
\includegraphics{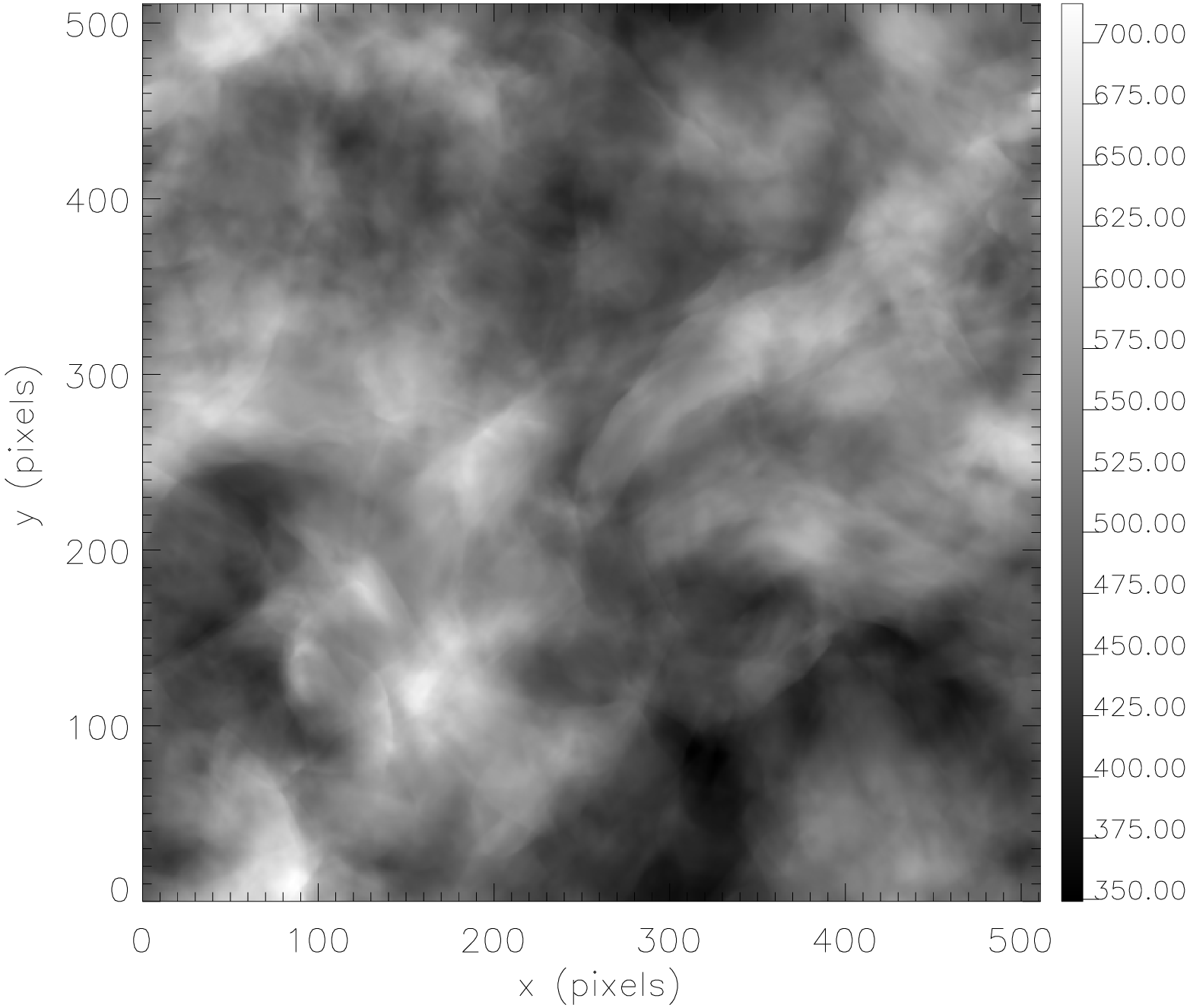}
}
\resizebox{\hsize}{!}{
\includegraphics{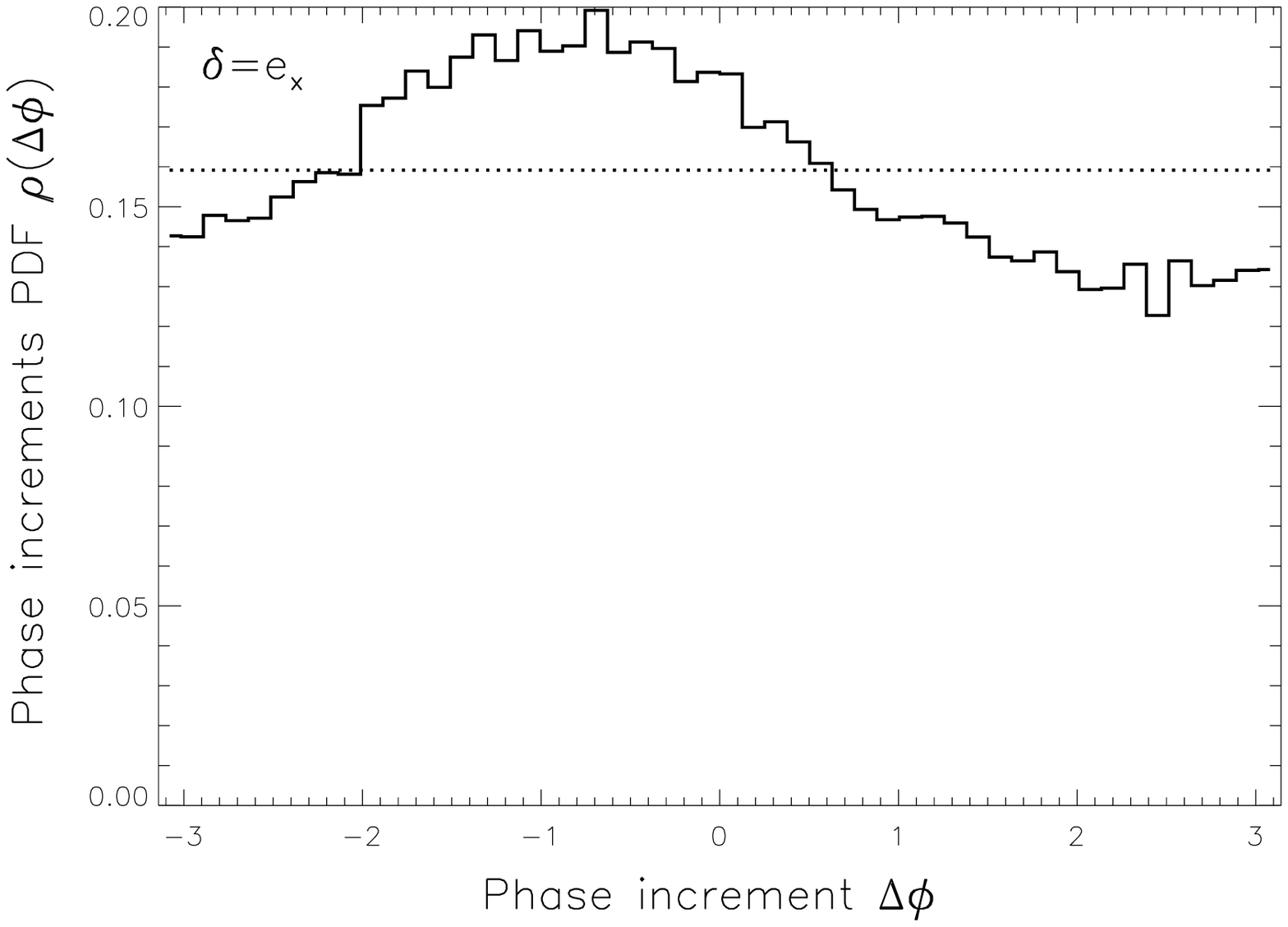}
}
\caption{{\it Top} : Column density of a 512$^3$ weakly compressible hydrodynamical turbulence simulation obtained by Porter {\it et al.} \cite{porter94}, used here as a model brightness distribution for phase structure analysis. {\it Bottom} : Histogram of phase increments for this field, with $\boldsymbol{\delta}=\boldsymbol{e}_x$ (unit vector along the $k_x$ axis in Fourier space) and $n=50$. The dotted line represents the uniform distribution.}
\label{fig_4}
\end{figure}

In a field for which Fourier phases are uncorrelated, such as fractional Brownian motions (fBm)\footnote{These are random fields characterized by a power-law power spectrum and random phases.} \cite{stutzki98}, phase increments are uniformly distributed over $[-\pi,\pi]$, for any lag vector $\boldsymbol{\delta}$. At the other end of the spectrum is the case of a single point source, for which the PDF of phase increments is a delta function. In between those extremes, the PDF of phase increments presents a single wavelike oscillation (See Fig.~\ref{fig_4} for an example), which may be seen as a signature of phase structure. 

A quantitative measure of the distribution's departure from uniformity is phase entropy \cite{polygiannakis95},
\begin{equation*}
\mathcal{S}(\boldsymbol{\delta})=-\int\nolimits_{-\pi}^{\pi} \rho\left(\Delta_{\boldsymbol{\delta}}\phi\right)\ln{\left[\rho\left(\Delta_{\boldsymbol{\delta}}\phi\right)\right]}\mathrm{d}\Delta_{\boldsymbol{\delta}}\phi,
\end{equation*}
which reaches its maximum value $\mathcal{S}_0=\ln{(2\pi)}$ for fBms. It is therefore convenient to consider the positive quantity $\mathcal{Q}(\boldsymbol{\delta})=\mathcal{S}_0-\mathcal{S}(\boldsymbol{\delta})$, which we dub \emph{phase structure quantity}, and which may be directly computed on the histograms of phase increments.

\section{Phase structure quantity in practice}
\label{sec_psqip}

For a finite-sized image, histograms of phase increments do not perfectly sample the underlying PDFs. Phase structure quantities $\mathcal{Q}$ associated with these distributions should be distinguished from those $\tilde{\mathcal{Q}}$ found by numerical integration of the histograms\footnote{To give an idea, for the histogram shown on Fig.~\ref{fig_4}, we have $\tilde{\mathcal{Q}}(\boldsymbol{e}_x)=9.8\times 10^{-3}$.}, which depend on the number $p$ of available increments and the number $n$ of bins. The difference between the two can become significant for $\mathcal{Q} \ll 1$.

In particular, finding $\tilde{\mathcal{Q}} \neq 0$ does not guarantee that phase structure is present in a given field ($\mathcal{Q} \neq 0$), so that the detectability of phase structure depends on the threshold of $\tilde{\mathcal{Q}}$ above which there is a given probability (say 0.99) that an image deviates significantly from a ``structureless'' field. 

The procedure is described in \cite{levrier2006} and is largely based on results from \cite{castellan2000}. In short, the end result is that the threshold of $\tilde{\mathcal{Q}}$ depends on $n$ and $p$, and may be found using well-known $\chi^2$ statistics. 

The influence of $n$ and $p$ on the reliability of $\tilde{\mathcal{Q}}$ may also be studied numerically, using fractional Brownian motions. Unsurprisingly, while $\mathcal{Q}=0$ for these, $\tilde{\mathcal{Q}}$ increases as the size of the image decreases, and as the number of bins increases.

\section{Application to interferometric observations}
\label{sec_atio}

In the ideal case, interferometers sample the Fourier transform of observed brightness distributions, and allow direct measurement of phase increments. Since this can be done as the Earth rotates, we may look for the minimum observing time required to detect a significant phase structure in the data. To focus on the problem of statistical estimation described in the previous section, we shall not consider primary beam attenuation nor regridding issues. These simplifications are discussed in more detail in \cite{levrier2006}. 

To estimate the ability of ALMA to detect and measure phase structure, we proceed as follows: A model brightness distribution is taken as input to a simple interferometer simulator, which is based on the characteristics of ALMA and uses the array configurations optimized by Boone \cite{boone2001}. The instrument tracks the source as long as it remains above a minimum elevation of 10$^\circ$. The output maps, for which no deconvolution is performed, yield values of $\tilde{\mathcal{Q}}$ as a function of integration time, with $\boldsymbol{\delta}$ and $n$ fixed.

The model brightness distributions used are the one of Fig.~\ref{fig_4}, and a field with the same power spectrum, but with random phases. For comparison, we have also considered configurations taken from current arrays, such as the Plateau de Bure (PdB) and the VLA, fictitiously located at the same geographical coordinates as ALMA, and observing the same source.

\begin{figure}[htbp]
\resizebox{\hsize}{!}{
\includegraphics{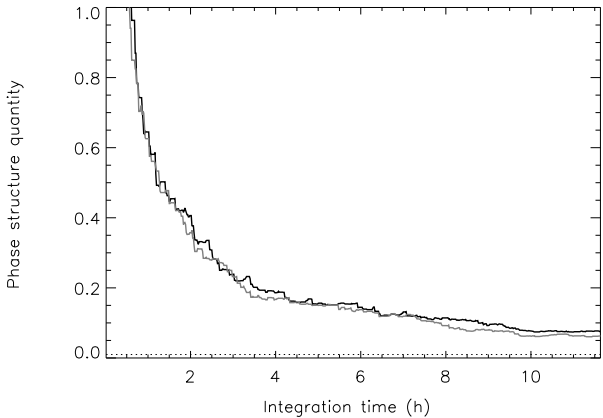}
}
\caption{Evolution of measured $\tilde{\mathcal{Q}}(\boldsymbol{e}_x)$ with integration time, for the B configuration of the Plateau de Bure. The black solid line corresponds to the turbulent brightness distribution, and the grey solid line to the random-phase brightness distribution. The dotted line represents $\tilde{\mathcal{Q}}(\boldsymbol{e}_x)$ for the complete turbulent brightness distribution, and the dashed line represents the evolution of the theoretical upper limit (lying above the plotted range here).}
\label{fig_n4}
\end{figure}

\begin{figure}[htbp]
\resizebox{\hsize}{!}{
\includegraphics{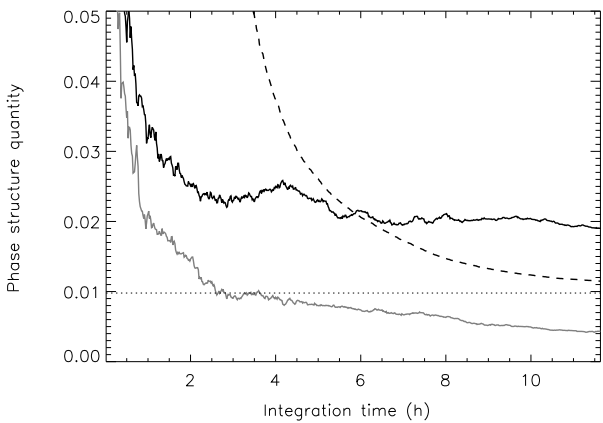}
}
\caption{Same as Fig.~\ref{fig_n4}, but for the D configuration of the VLA.}
\label{fig_n5}
\end{figure}

\begin{figure}[htbp]
\resizebox{\hsize}{!}{
\includegraphics{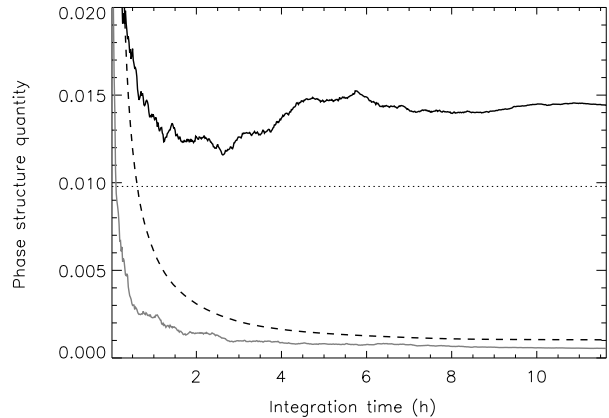}
}
\caption{Same as Fig.~\ref{fig_n4}, but for the E configuration of ALMA.}
\label{fig_n1}
\end{figure}

As the observation is carried out, more and more Fourier phases are measured and $p$ increases. The question is whether this allows to bring down the upper limit discussed in section \ref{sec_psqip}, below the measured phase structure quantities, to ensure positive detection. The results are summarized on Figures~\ref{fig_n4} to~\ref{fig_n1}, which show the evolution of $\tilde{\mathcal{Q}}(\boldsymbol{e}_x)$ as a function of integration time.
Fig.~\ref{fig_n4} shows that the number of phase increments measured by the Plateau de Bure in its B configuration is insufficient to detect phase structure, as the curves for turbulent and random-phase brightness distributions are indistinguishable from one another. The same conclusion prevails for other configurations of this instrument and other lag vectors. 
On the contrary, Fig.~\ref{fig_n5} shows that the VLA allows such a detection, since the measured $\tilde{\mathcal{Q}}$ becomes larger than the theoretical upper limit, after about 6 hours of integration. Long before that, however, we get a hint that phase structure is present in the field, since the curves for both model brightness distributions go apart after less than twenty minutes. This diagnosis can be performed in real time by drawing random phases for the visibilities as they are measured.
ALMA gives even better results (Fig.~\ref{fig_n1}). In its E configuration and in our case, a short integration time of about twenty minutes is enough to conclude on the presence of phase structure. However, the final value of $\tilde{\mathcal{Q}}$ obtained is not equal to the phase structure quantity measured on the model brightness distribution. This is due to the fact that only 24\% of the $512 \times 512$ Fourier phases are measured by this configuration. 

\begin{figure}[htbp]
\resizebox{\hsize}{!}{
\includegraphics{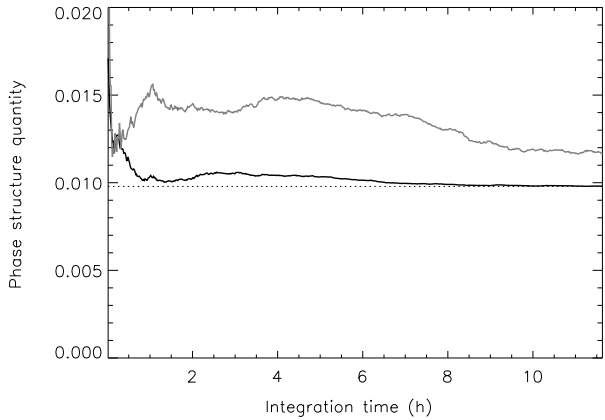}
}
\caption{Evolution of measured $\tilde{\mathcal{Q}}(\boldsymbol{e}_x)$ with integration time for an observation using all configurations of the instrument in turn. The black solid line corresponds to the six configurations of ALMA, and the grey solid line to the four configurations of the VLA. The dotted line represents the value of $\tilde{\mathcal{Q}}(\boldsymbol{e}_x)$ for the whole field.}
\label{fig_n7}
\end{figure}

Using more extended configurations, one should be able to measure the Fourier components lying outside the radius covered by the E configuration, and therefore hope to recover the correct value of the phase structure quantity by combining visibilities from multiple configurations. Fig.~\ref{fig_n7} shows the evolution of the measured $\tilde{\mathcal{Q}}(\boldsymbol{e}_x)$ with integration time, using this approach\footnote{The integration time $\tau$ is to be understood per configuration, and the total time of integration is $N_{\textrm{configurations}}\times\tau$.}. It appears that the Fourier plane coverage achieved by ALMA will allow measurement of the actual value of the phase structure quantity for the observed field, while the VLA fails.

Finally, to assess whether atmospheric phase noise would prevent detection of phase structure, we introduced a mask giving the refractivity field above the instrument. We assumed this mask to be a 200-m thick layer of frozen Kolmogorov turbulence being transported along the east-west direction at 2 m.s$^{-1}$, and normalized it so that the rms phase noise $\sigma_0$ for a pair of antennae observing the zenith and separated by a baseline $d=100$ m should be one of a few specific values, namely 15$^\circ$, 45$^\circ$ and 90$^\circ$. According to \cite{butler2001} and using the scaling relation given by \cite{lay97a}, noise levels at Chajnantor vary typically from $\sigma_0 \sim 14^{\circ}$ to $\sigma_0 \sim 57^{\circ}$. 

\begin{figure}[htbp]
\resizebox{\hsize}{!}{
\includegraphics{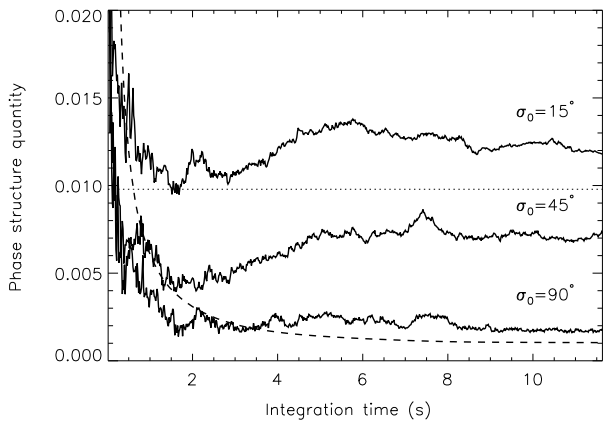}
}
\caption{Evolution of measured $\tilde{\mathcal{Q}}(\boldsymbol{e}_x)$ with integration time in the presence of atmospheric phase noise (solid lines, with $\sigma_0$ specified next to each curve). The array used is the E configuration of ALMA. The dotted line represents $\tilde{\mathcal{Q}}(\boldsymbol{e}_x)$ for the whole field, and the dashed line shows the theoretical upper limit.}
\label{fig_n6}
\end{figure}

Integration of the refractivity field along the different lines of sight for each antenna as the observation is performed yields phase delays, which are then correlated to obtain the atmospheric phase noise for each pair of antennae, at all times. Fig.~\ref{fig_n6} shows the evolution of the measured $\tilde{\mathcal{Q}}(\boldsymbol{e}_x)$ for ALMA in its E configuration. 

It appears that in this case, the presence of phase structure can be easily detected in the presence of a fair amount of atmospheric phase noise. Indeed, even a rms phase fluctuation of $\sigma_0=90^{\circ}$ is insufficient to bring the measured phase structure quantity below the upper limit. Consequently, phase structure will undoubtedly be detected by ALMA without any phase correction, although the use of dedicated water vapor radiometers, as is planned, should allow for an effective decrease of the atmospheric phase noise by a substantial factor \cite{lay97b}, making it possible to actually measure the phase structure quantity for the observed field.

\section{Perspectives}
\label{sec_conc}

In the context of interferometry, a more elaborate use of phase information would be to keep track of the phase measured by each baseline as a function of time, and to compute phase increments along the baseline's track. This should reduce contamination by atmospheric phase noise, but would require a shift in the phase structure information formalism, since, in this approach, the lag vector $\boldsymbol{\delta}$ is no longer a control parameter, but a function of time and of the baseline.

Another possible extension of this work is the inclusion of the kinematic dimension, which is accessible through ALMA's high spectral resolution receivers. It may well be that phase analysis applied to individual channel maps should prove a valuable tool for assessing the three-dimensional structure of velocity fields.

\end{document}